\documentclass[10pt,twocolumn,letterpaper]{article}

\usepackage[margin=0.9in]{geometry}
\usepackage{times}
\usepackage{graphicx}
\usepackage{amsmath}
\usepackage{amssymb}

\usepackage{subcaption}
\usepackage{multirow}

\graphicspath{{figures/}}

\begin{document}

\title{Demoir\'eing of Camera-Captured Screen Images Using Deep
  Convolutional Neural Network}

\author{Bolin Liu\\
	McMaster University\\
	{\tt\small liub30@mcmaster.ca}
	\and
	Xiao Shu\\
	Shanghai Jiao Tong University\\
	{\tt\small shux@sjtu.edu.cn}
	\and
	Xiaolin Wu\\
	Shanghai Jiao Tong University\\
	{\tt\small xwu510@sjtu.edu.cn}
}

\date{}

\maketitle
\thispagestyle{empty}

\begin{abstract}
  Taking photos of optoelectronic displays is a direct and spontaneous
  way of transferring data and keeping records, which is widely
  practiced.  However, due to the analog signal interference between
  the pixel grids of the display screen and camera sensor array,
  objectionable moir\'e (alias) patterns appear in captured screen
  images.  As the moir\'e patterns are structured and highly variant,
  they are difficult to be completely removed without affecting the
  underneath latent image.  In this paper, we propose an approach of
  deep convolutional neural network for demoir\'eing screen photos.
  The proposed DCNN consists of a coarse-scale network and a
  fine-scale network.  In the coarse-scale network, the input image is
  first downsampled and then processed by stacked residual blocks to
  remove the moir\'e artifacts.  After that, the fine-scale network
  upsamples the demoir\'ed low-resolution image back to the original
  resolution.  Extensive experimental results have demonstrated that
  the proposed technique can efficiently remove the moir\'e patterns
  for camera acquired screen images; the new technique outperforms the
  existing ones.
\end{abstract}

\section{Introduction}
\label{sec:introduction}

\begin{figure}
  \centering
  \begin{subfigure}{0.45\linewidth}
    \centering
    \includegraphics[width=0.9\linewidth]{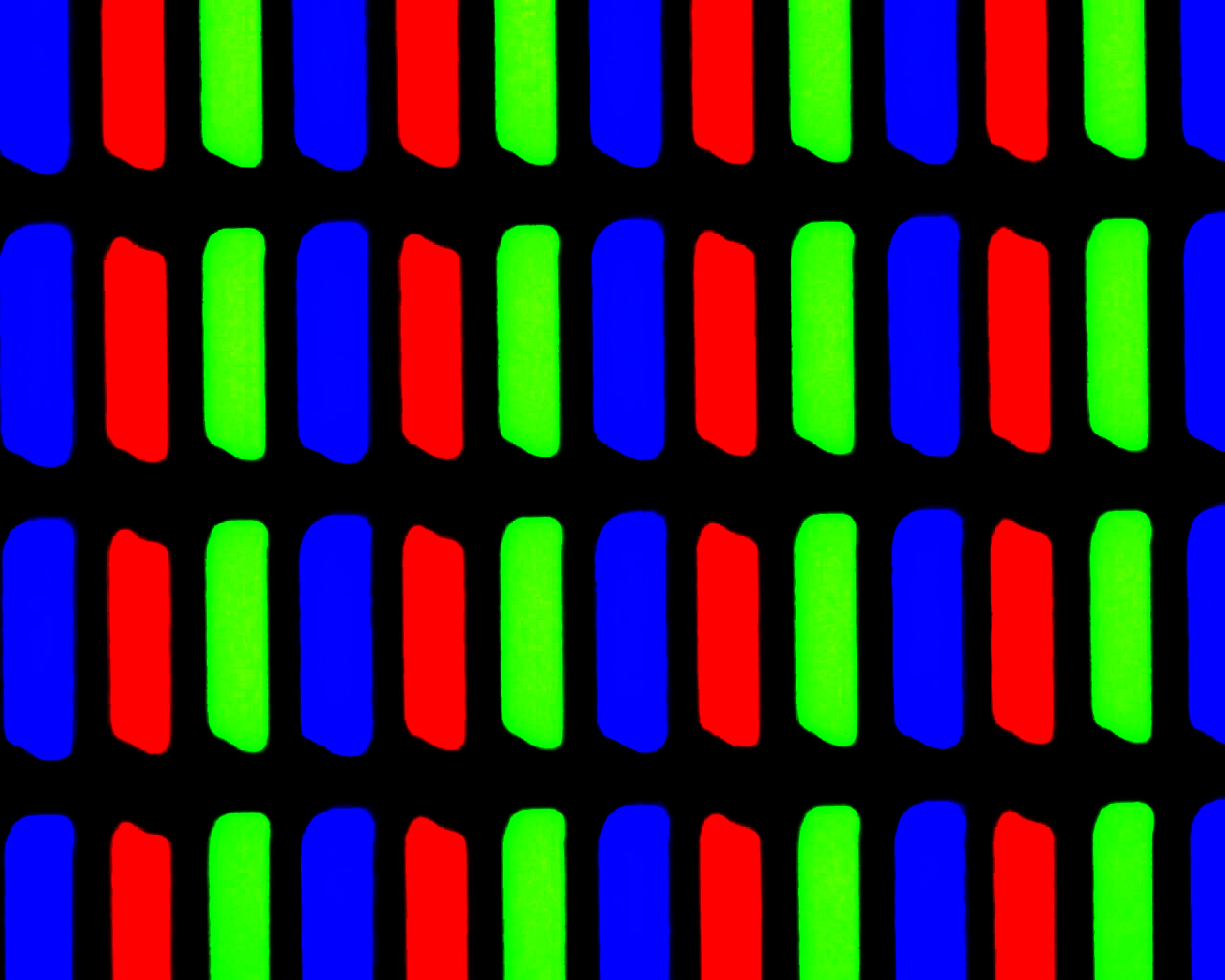}
    \caption{LCD subpixel structure}
    \label{fig:synth_lcd}
  \end{subfigure}
  \begin{subfigure}{0.45\linewidth}
    \centering
    \includegraphics[width=0.9\linewidth]{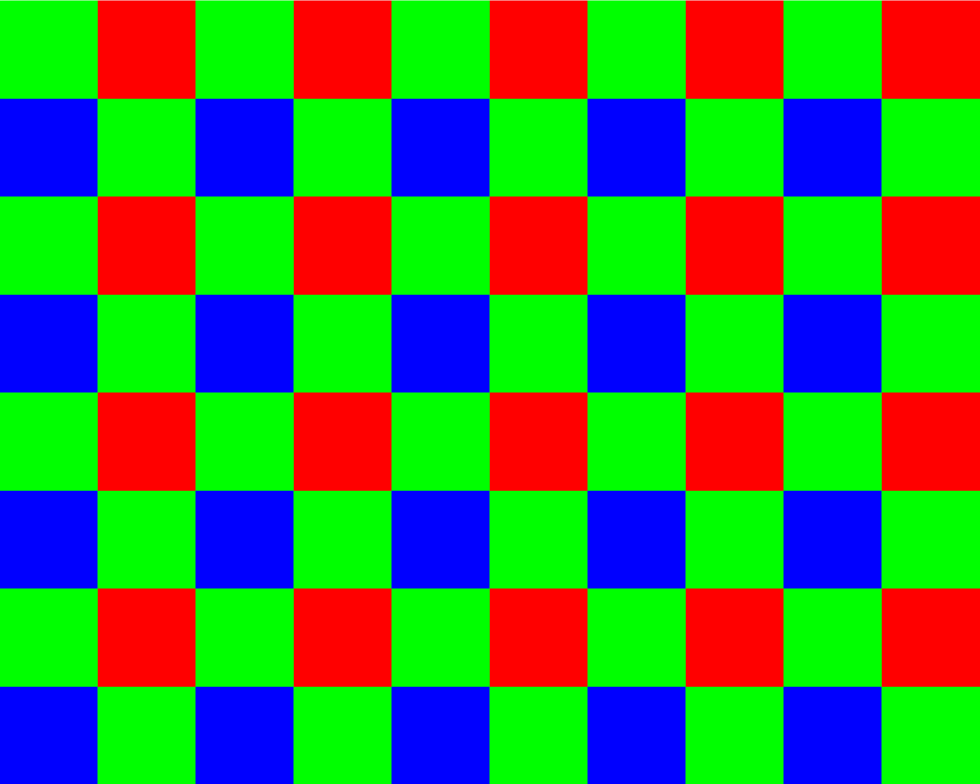}
    \caption{Camera Bayer CFA}
    \label{fig:synth_cam}
  \end{subfigure}
  \caption{Image displayed on a LCD or captured by a camera is
    spatially resampled in order to achieve color effects.}
  \label{fig:synth}
\end{figure}

Capturing screen-displayed contents by cameras has become for many a
spontaneous and convenient way of exchanging information and keeping
records across different media platforms.  This user behavior is quite
natural due to the ubiquity and wide uses of digital cameras and
information displays of all types, particularly those integrated into
portable personal devices such as smartphones and tablets.  In terms
of multimedia interface, the alternative of issuing print-screen
command, saving the resulting image file and having it emailed is
cumbersome and less intuitive.  In some situations digitally saving
the displayed image is not even possible, taking public displays for
example.  However, the analog means of camera shooting screens can
severely degrade the image quality.  As both the display screen (as
shown in Fig.~\ref{fig:synth_lcd}) and camera sensor array (as shown
in Fig.~\ref{fig:synth_cam}) need to resample an image in order to
achieve color effects, the interference between the sampling grids of
the display and camera often generates objectionable moir\'e
(aliasing) patterns.


\begin{figure}
  \centering
  \begin{subfigure}[t]{0.32\linewidth}
    \centering
    \includegraphics[width=\linewidth]{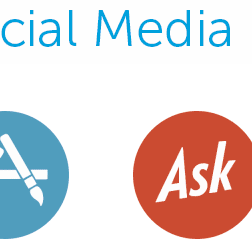}
    \caption{Original image}
  \end{subfigure}
  \begin{subfigure}[t]{0.32\linewidth}
    \centering
    \includegraphics[width=\linewidth]{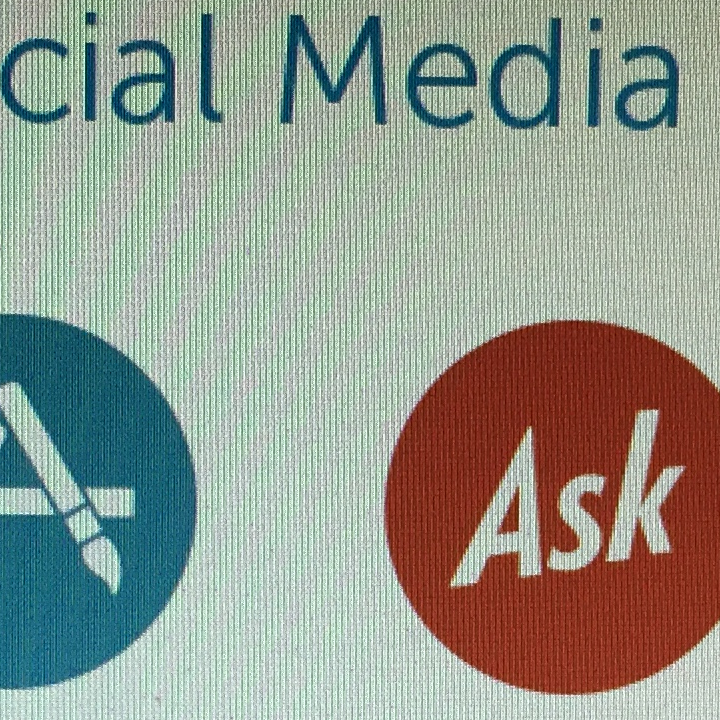}
    \caption{Camera-captured screen image}
  \end{subfigure}
  \begin{subfigure}[t]{0.32\linewidth}
    \centering
    \includegraphics[width=\linewidth]{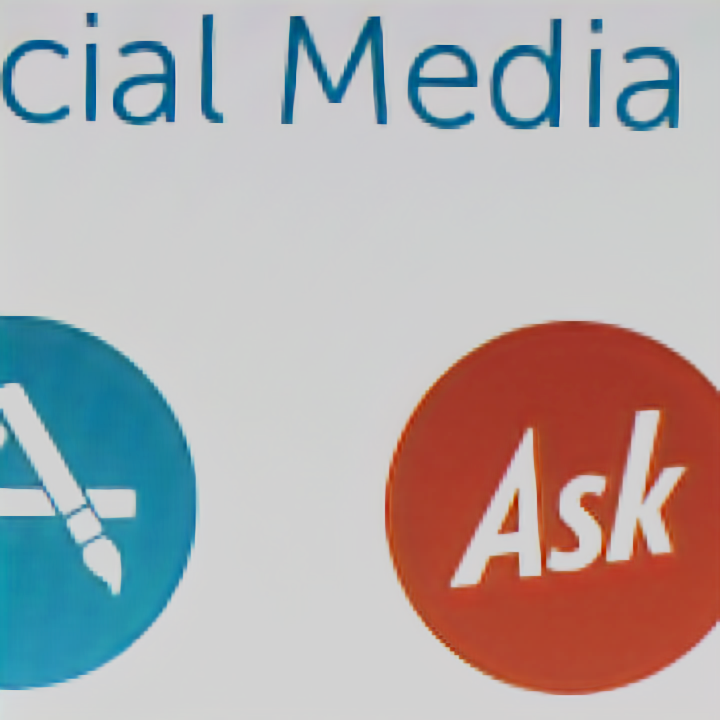}
    \caption{Moir\'e artifacts removed image}
  \end{subfigure}
  \caption{The proposed technique is designed to remove moir\'e
    artifacts in camera-captured screen image and to recover the
    original image intended to display on the screen.}
  \label{fig:purpose}
\end{figure}

Despite the commonness and annoyance of the problem, little work has
been done on the reduction of moir\'e artifacts in camera-captured
screen images.  In attempt to suppress moir\'e patterns, a camera
often has a layer of optical low-pass (anti-aliasing) filter placed in
front of the sensor arrays.  But unless compromising the sharpness of
an image, the optical approach has limited effectiveness against
strong moir\'e patterns commonly found in camera-captured screen
images.  Another approach is to use adaptive digital filtering
\cite{yang2017demoireing}.  While this method shows some improvement
over the optical approach, its results are still far from
satisfactory.  After all, it is very difficult, if not impossible, for
conventional signal processing methods to remove aliases like moir\'e
patterns completely, as suggested by the Nyquist-Shannon sampling
theorem \cite{shannon1949communication}.

In this work, we tackle the problem of restoring camera-captured
screen images against moir\'e artifacts using neural network
techniques, as shown in Fig.~\ref{fig:purpose}.  Machine learning
methods, deep convolutional neural networks (DCNN) in particular, hold
the promise to satisfactorily solve the demoir\'eing problem, as they
can exploit and benefit from the distinctive statistics of
moir\'e-free images and moir\'e patterns learned from suitable
training data.

In the development of our DCNN technique for demoir\'eing, we face the
challenge of obtaining exactly matched pairs of real
moir\'e-interfered screenshot and its corresponding clean image for
training; it is very difficult to generate moir\'e-free screenshot and
to spatially align the pair of images perfectly.  Thus, instead of
relying solely on real captured images, we carefully model the
formation of moir\'e patterns during the capture of an LCD screen
using a Bayer color filter array (CFA) camera, and generate a large
number of synthetic training images with realistic moir\'e patterns
from original digital images.  While data synthesis brings convenience
of building a large training set quickly, synthetic data may still
deviate from real camera-captured screen images in certain aspects
such as camera shake, chromatic aberration, unclean screen surface,
reflection, etc.  To improve the robustness of the proposed technique
in realistic settings, we introduce a novel two-stage training
procedure, which pretrains the neural network with synthetic data and
then retrains it using real images and the results from the first
stage.  Another contribution of this work is the use of a multi-scale
neural network architecture.  As moir\'e patterns exist and exhibit
vastly different characteristics in different scales, we find that
removing moir\'e artifacts gradually, from coarse to fine, can greatly
improve the effectiveness of the proposed DCNN technique.

The remainder of the paper is organized as follows.  Section
\ref{sec:related} provides an overview of related work in the
literature.  In Section \ref{sec:data}, we introduce our method for
synthesizing training data, and in Section \ref{sec:algorithm}, we
present in detail our proposed method.  Section \ref{sec:result} shows
the experiment evaluations of the proposed method with both synthetic
and real-world images.  Finally, Section \ref{sec:conclusion}
concludes.

\section{Related Work}
\label{sec:related}

To the best of our knowledge, the only existing published work on the
problem of demoir\'eing for camera-captured screen images is a
conventional image processing technique, called layer decomposition on
polyphase components (LDPC) \cite{yang2017demoireing}.  The LDPC
technique has three major steps: it first subsamples the input image
into four polyphase components; next, for each component, it separates
the moir\'e interference layer based on a patch based Gaussian mixture
model (GMM) prior; at last, the technique recombines the four filtered
polyphase components into one full resolution image as the output.
Although LDPC can remove some small moir\'e artifacts, it tends to
over-smooth the image detail and it cannot handle large-scale moir\'e
patterns, such as color stripes, due to its very small patch size.

The authors of LDPC also proposed a demoir\'eing technique for
textured images, i.e., images of objects with some fine grid patterns,
like fabric \cite{yang2017textured}.  This technique removes moir\'e
artifacts in green channel using signal decomposition and reconstructs
the red and blue channels using guided image filter with the cleaned
green channel as the guide.  Another related demoir\'eing problem is
the removal of moir\'e patterns for scanned images, where the artifacts
have net-like structures.  The techniques designed for solving this
problem are commonly referred as descreening.  One of the early
descreening algorithms is \cite{luo1998robust}, which employs wavelet
domain filtering technique to remove moir\'e patterns.  Siddiqui et
al. proposed a non-iterative and non-linear descreening filter based
on resolution synthesis-based denoising \cite{siddiqui2007training}.
Shou and Lin presented a technique that employs cellular neural
network based texture classification for moir\'e pattern screening
\cite{shou2004image}.

The demoir\'eing problem can also be categorized as one of image
restoration.  There have been quite a few inspiring image restoration
techniques proposed in the past decade, such as non-local similarity
based techniques \cite{dabov2007image}, nuclear norm minimization
based techniques \cite{gu2014weighted}, dictionary learning based
techniques \cite{elad2006image, chatterjee2009clustering,
  wang2015learning}.  Recently, deep neural network based techniques
have demonstrated their great strength in many different fields and
achieved the state-of-the-art for various image restoration problems.
For example, stacked denoising auto-encoder achieves good results for
denoising corrupted images \cite{vincent2010stacked}; multi-layer
perceptron (MLP) can be used as a denoising framework
\cite{burger2012image}.  For more general image restoration tasks, Mao
proposed RED-Net, a very deep convolutional neural network with skip
connections \cite{mao2016image}.  Another example of general image
restoration network is DnCNN.  By combining batch normalization and
residual learning method, DnCNN can tackle various problems, such as
Gaussian denoising, single image super-resolution and JPEG deblocking
\cite{zhang2017beyond}.  By retraining these DCNN methods with
moir\'e-interfered images, they can be repurposed for solving
demoir\'eing problem as well.

\section{Preparation of Training Data}
\label{sec:data}

The basic idea of the proposed demoir\'eing technique is to map a
moir\'e pattern tainted screen photo to an artifact-free image using
an end-to-end neural network trained with a large number of such image
pairs.  The effectiveness of our technique, or any machine learning
approaches, greatly relies on the availability of a representative and
sufficiently large set of training data.  In this section, we discuss
the methods for preparing the training images for our technique.

To help the proposed DCNN technique identify the moir\'e artifacts
accurately in real-world scenarios, ideally, the training process
should only use real photographs of a screen and the corresponding
original digital images displayed on it.  While obtaining such a pair
of images is easy, perfectly aligning them spatially, a necessary
condition for preventing mismatched edges being misidentified as
moir\'e patterns, is difficult to achieve.  As many common imaging
problems in real photos, such as lens distortion and non-uniform
camera shake, adversely affect the accuracy of image alignment, it is
challenging to build a sufficiently large and high quality training
set using real photos.

Considering the drawbacks of using real photos, we employ synthetic
screenshot images with realistic moir\'e patterns for training
instead.  The input images of the synthesizer, which are collected by
using the print-screen command from computers running Microsoft
Windows, cover various types of content, such as dialog box, text, web
page, graphics, natural images, etc.  To accurately simulate the
formation of moir\'e patterns, we follow truthfully the process of
image display on an LCD and the pipeline of optical image capture and
digital processing on a camera.  The whole simulation procedure can be
summarized in the following steps.
\begin{enumerate}
\item Resample the input image into a mosaic of RGB subpixels as in
  Fig.~\ref{fig:synth_lcd} to simulate the image displayed on the LCD;
\item Apply a projective transformation with a certain degree of
  randomness on the image to simulate different relative positions and
  orientations of the display and camera;
\item Use radial distortion function to simulate lens distortion;
\item Apply a flat top Guassian filter to simulate anti-aliasing
  filter;
\item Resample the image using Bayer CFA as in
  Fig.~\ref{fig:synth_cam} to simulate the raw reading of the camera
  sensor;
\item Add Guassian noise to simulate sensor noise;
\item Apply demosaicing;
\item Apply denoising filter;
\item Compress the image using JPEG to add compression noise;
\item Output the decompressed image as the synthetic image with
  moir\'e patterns
\end{enumerate}
The corresponding groundtruth clean image is generated from the
original image using the same projective transformation and lens
distortion function.  In addition, the groundtruth image is also
scaled to match the same size of the synthetic camera-captured screen
image.

\section{Proposed Algorithm}
\label{sec:algorithm}

\begin{figure}
  \centering
  \begin{subfigure}[t]{0.45\linewidth}
    \centering
    \includegraphics[width=0.9\linewidth]{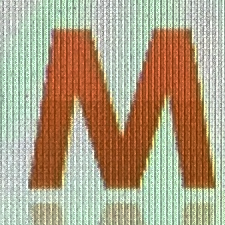}
    \caption{Fine scale}
  \end{subfigure}
  \begin{subfigure}[t]{0.45\linewidth}
    \centering
    \includegraphics[width=0.9\linewidth]{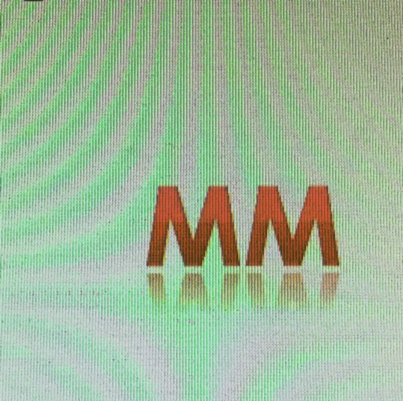}
    \caption{Coarse scale}
  \end{subfigure}
  \caption{Moir\'e patterns in camera-captured screen images appear
    very differently in different scales.}
  \label{fig:moire}
\end{figure}

Unlike the degradations in many other image restoration problems,
moir\'e patterns are signal-dependent and structured; they often can
only be distinguished from true image features when we examine both
the big picture and fine details.  Additionally, the moir\'e patterns
in a camera-captured screen image appear very differently in different
scales.  As exemplified in Fig.~\ref{fig:moire}, the moir\'e patterns
are thin vertical stripes in fine scale; while they look like curved
color bands in coarse scale.  Due to these complex characteristics of
moir\'e patterns, most existing DCNN architectures are unsuitable and
ineffective for the demoir\'eing task.

\begin{figure*}
  \centering
  \includegraphics[width=\linewidth]{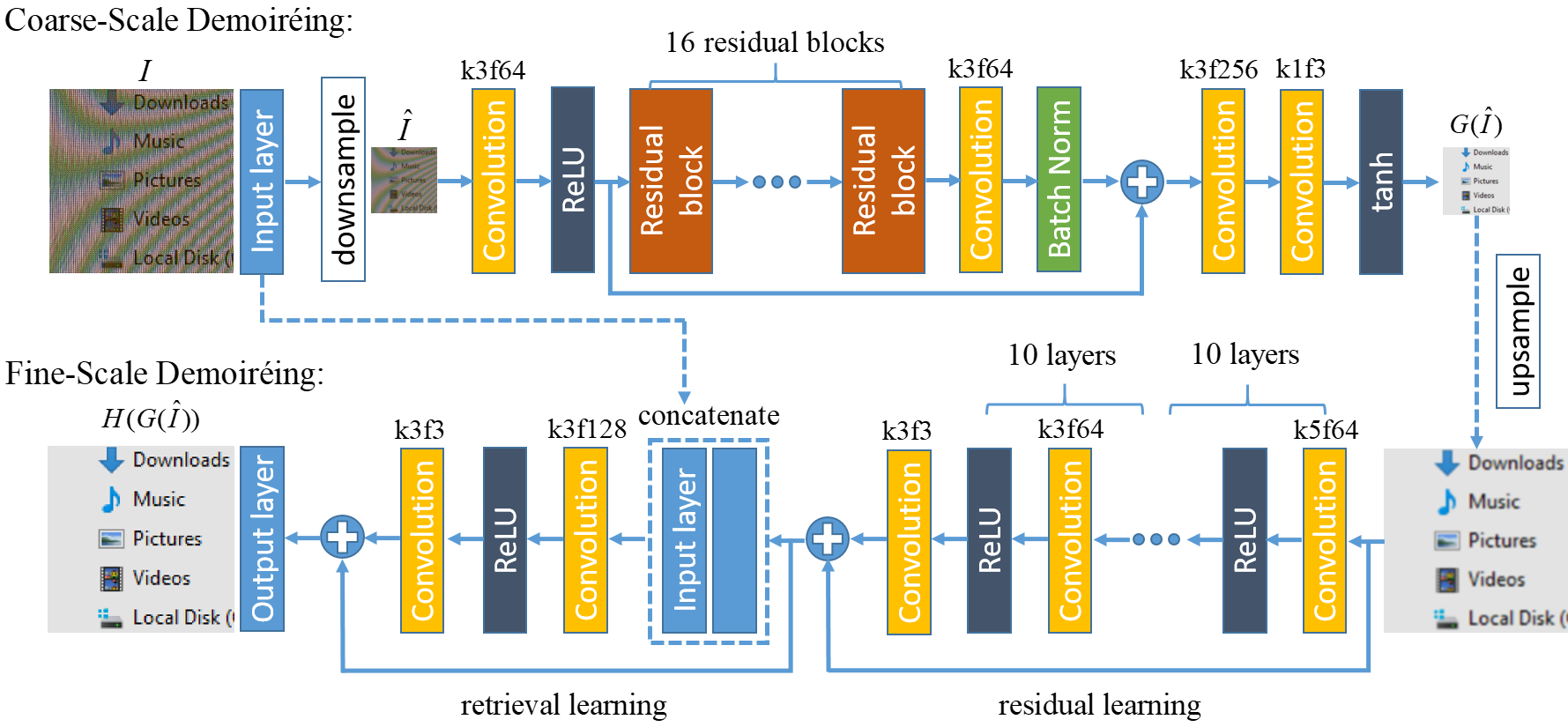}
  \caption{The architecture of the proposed neural network.  The
    kernel size $k$ and the number of feature maps $f$ are marked
    above each convolutional layer.}
  \label{fig:allnets}
\end{figure*}

To deal with these aforementioned difficulties, the proposed
demoir\'eing technique adopts a multi-scale strategy, as sketched in
Fig.~\ref{fig:allnets}.  The basic idea is as follows.  For an input
camera-captured screen image $I$, the proposed technique first blurs
and downsamples $I$ using a Gaussian kernel, yielding image $\hat{I}$.
Next, the technique employs a generator neural network $G$ trained
with downsampled screen images to reduce the moir\'e patterns in
$\hat{I}$.  The resulting image $G(\hat{I})$ is then upsampled back to
the original scale and sent to another network $H$ along with the
original input image $I$ for fine-scale moir\'e pattern removal.  The
following are the details of the proposed technique.

\subsection{Coarse-Scale Demoir\'eing}

\begin{figure}
  \centering
  \includegraphics[width=0.9\linewidth]{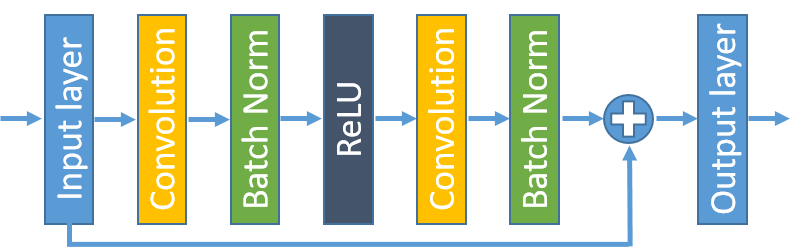}
  \caption{The structure of a residual block.}
  \label{fig:resblock}
\end{figure}

The coarse-scale demoir\'eing is achieved by using a generator network
$G$ to map a downsampled camera-captured screen image $\hat{I}$ to its
artifact-free counter part $\hat{J}$.  Inspired by the work of Ledig
et al. \cite{ledig2016photo}, we employ stacked residual blocks as the
foundation of the architecture of network $G$
\cite{gross2016training}.  Illustrated in Fig.~\ref{fig:resblock} is
the structure of a residual block, which consists of two convolutional
layers with $3 \times 3$ kernels and 64 feature maps, followed by
batch normalization layers \cite{ioffe2015batch} and rectified linear
units (ReLU) \cite{nair2010rectified}.  As shown in
Fig.~\ref{fig:allnets}, there are 16 such residual blocks in total in
the network, and at the end, there are a skip connection to add back
some details and a hyperbolic tangent function $\tanh(\cdot)$ as the
final nonlinear operation.

With the synthetic screen image set discussed in the previous section,
we can use supervised learning to train a generator network $G'$ using
downsampled versions of the synthetic moir\'e-interfered and
artifact-free image pairs.  The loss function of network $G'$ for the
training process is a mean squared error (MSE) function defined as
follows,
\begin{align}
  L'_{\ell_2} = \frac{1}{MN}||G'(\hat{I}) - \hat{J}||_2^2,
\end{align}
where the $M$ and $N$ are the height and width of downsampled image
$\hat{I}$, respectively, and $\hat{J}$ is the downsampled groundtruth
image.

\subsection{Retraining with Real Images}

The synthetic data trained network $G'$ performs well against
artificially generated moir\'e patterns in coarse scale.  However,
since the data synthesizer cannot cover every possible characteristics
of real camera-captured screen images, the network $G'$ may fail to
identify all the moir\'e patterns fully in a real image, leaving
traces of artifacts.

\begin{figure}
  \centering
  \includegraphics[width=\linewidth]{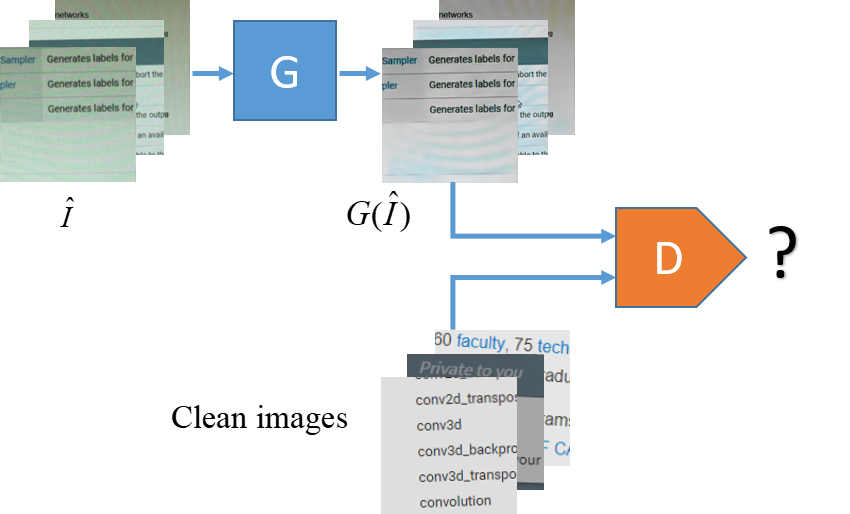}
  \caption{The workflow of proposed GAN, where $G$ represents
    generator and $D$ represents discriminator.}
  \label{fig:gan_sketch}
\end{figure}

To improve the robustness of the proposed technique against real
images, we integrate idea of generative adversarial network (GAN) in
our architecture \cite{goodfellow2014generative}, as shown in
Fig.~\ref{fig:gan_sketch}.  In GAN, a discriminative network $D$ is
jointly trained with the generative network $G$ for discriminating the
output images of $G$ against a set of unrelated original clean images.
This process guides $G$ to return images that are statistically
similar to artifact-free images.  As the training for GAN is not
necessarily paired, we can train the network using real images even if
the corresponding groundtruth images are unavailable.

Following the idea of Goodfellow et al., we set the discriminative
network $D$ to solve the following minimax problem:
\begin{align}
  \min_G \max_D \left\{ \mathbb{E}_{\tilde{J}} \left[ \log D(\tilde{J}) \right] +
    \mathbb{E}_{\hat{I}} \left[ \log(1-D(G(\hat{I}))) \right] \right\}.
\end{align}
In practice, for better gradient behavior, we minimize $-\log
D(G(\hat{I}))$ instead of $\log(1-D(G(\hat{I})))$, as proposed in
\cite{goodfellow2014generative}.  This introduces an adversarial term
in the loss function of the generator network $G$:
\begin{align}
  L_\oslash = -\log D(G(\hat{I})) .
\end{align}
In competition against generator network $G$, the loss function for
training discriminative network $D$ is the binary cross entropy:
\begin{align}
  L_D = - \left[ \log(D(\tilde{J}))+\log(1-D(G(\hat{I}))) \right].
\label{loss_D}
\end{align}
Minimizing $L_\oslash$ drives network $G$ to produce images that
network $D$ cannot distinguish from original artifact-free images.
Accompanying the evolution of $G$, minimizing $L_D$ increases the
discrimination power of network $D$.

However, optimizing for loss function $L_\oslash$ alone does not
guarantee that an output image $G(\hat{I})$ is similar to the input
moir\'e-interfered screenshot image $\hat{I}$; a fidelity term is
necessary for high quality restoration.  In the proposed technique,
instead of using the input image $\hat{I}$ to regulate $G(\hat{I})$,
we employ $G'(\hat{I})$, the result from the synthetic data trained
network $G'$, to formulate the following loss function.

\begin{align}
  L_{\ell_2} = \frac{1}{MN}||G(\hat{I}) - G'(\hat{I})||_2^2.
\end{align}

Finally, we combine the adversarial loss and fidelity loss when
optimizing the generative network $G$, namely,
\begin{equation}
  L_G = L_{\ell_2} + \lambda L_\oslash.
  \label{eq:joint training}
\end{equation}
where Lagrange multiplier $\lambda$ is a user given weight balancing
the two loss functions.  In all the presented experiments, $\lambda$
is set to $10^{-4}$ empirically.

\begin{figure}
  \centering
  \begin{subfigure}{0.32\linewidth}
    \includegraphics[width=\linewidth]{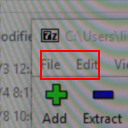}
    \includegraphics[width=\linewidth]{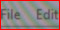}
    \\[1ex]
    \includegraphics[width=\linewidth]{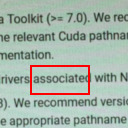}
    \includegraphics[width=\linewidth]{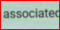}
    \caption{Original}
  \end{subfigure}
  \begin{subfigure}{0.32\linewidth}
    \includegraphics[width=\linewidth]{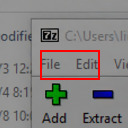}
    \includegraphics[width=\linewidth]{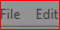}
    \\[1ex]
    \includegraphics[width=\linewidth]{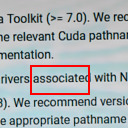}
    \includegraphics[width=\linewidth]{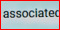}
    \caption{Result of $G'$}
  \end{subfigure}
  \begin{subfigure}{0.32\linewidth}
    \includegraphics[width=\linewidth]{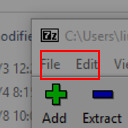}
    \includegraphics[width=\linewidth]{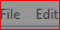}
    \\[1ex]
    \includegraphics[width=\linewidth]{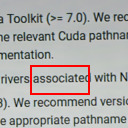}
    \includegraphics[width=\linewidth]{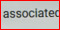}
    \caption{Result of $G$}
  \end{subfigure}

  \caption{The synthetic data trained network $G'$ works well for
    synthetic image (first row), but it is inferior to the real data
    retrained network $G$ for real images (second row).}
  \label{fig:exp_gan}
\end{figure}

As demonstrated in Fig.~\ref{fig:exp_gan}, the synthetic data trained
network $G'$ works well for synthetic camera-captured screen image,
but it fails to remove all the moir\'e patterns for real image.  In
comparison, the network $G$, which is retrained using real data,
removes the artifacts completely for both synthetic and real images.

\subsection{Fine-Scale Demoir\'eing}
\label{sec:superres}

At this stage, we have the high-resolution moir\'e-interfered input
image $I$ and the low-resolution but demoir\'ed image $G(\hat{I})$,
and we want to recover a high-resolution demoir\'ed image from these
two images.  The problem is akin to a super-resolution (SR) of
$G(\hat{I})$, except that extra information, a degraded version $I$ of
the original image, is available.

Intuitively, we can first upsample $G(\hat{I})$ to the original
resolution using SR and then merge the result with $I$ to produce the
final output.  Based on this idea, we design a fine-scale demoir\'eing
network $H$, as shown in the bottom half of Fig.~\ref{fig:allnets}.
The proposed network consists of two stages: residual learning and
retrieval learning.  In the residual learning stage, the proposed
network takes bicubic interpolated image as the input.  Then,
following the architecture of VDSR \cite{kim2016accurate}, the network
uses 20 cascaded convolutional layers with the ReLU activation to
recover the missing details.  In first 10 layers, the kernel size is
$5 \times 5$, and in next 10 layers, the kernel size is $3 \times 3$.
64 filters are used in each convolutional layer and a skip connection
is employed after the last residual layer.

In the retrieval stage, the proposed network concatenates the output
of residual learning stage with the original input $I$ and uses two
convolutional layers to refine the final result using $I$ and
upsampled $G(\hat{I})$.  With synthetic data, we employ the MSE as the
loss function of the fine-scale demoir\'eing network $H'$ as follows,
\begin{align}
  L_{H'} = \frac{1}{MN}||H'(G(\hat{I})) - J||_2^2.
\end{align}
where the $M$ and $N$ are the height and width of the input image $I$,
respectively, and $J$ is the groundtruth clean image.

\begin{figure}
  \centering
  \begin{subfigure}{0.32\linewidth}
    \includegraphics[width=\linewidth]{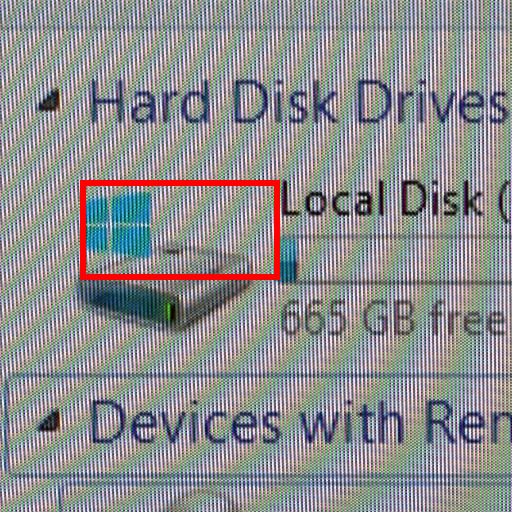}
    \includegraphics[width=\linewidth]{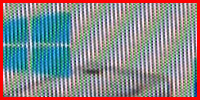}
    \\[1ex]
    \includegraphics[width=\linewidth]{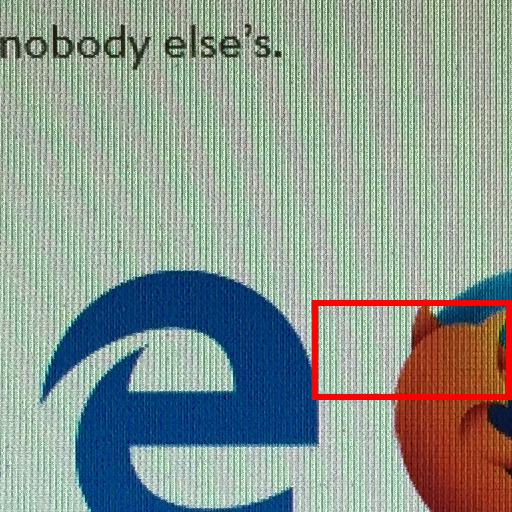}
    \includegraphics[width=\linewidth]{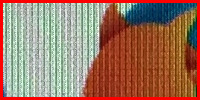}
    \caption{Original}
  \end{subfigure}
  \begin{subfigure}{0.32\linewidth}
    \includegraphics[width=\linewidth]{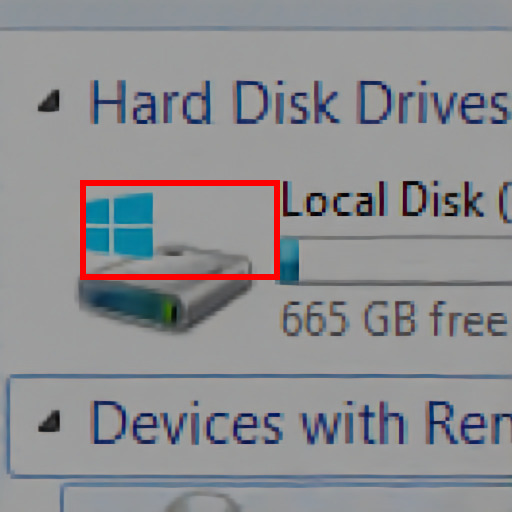}
    \includegraphics[width=\linewidth]{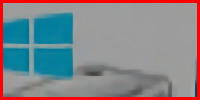}
    \\[1ex]
    \includegraphics[width=\linewidth]{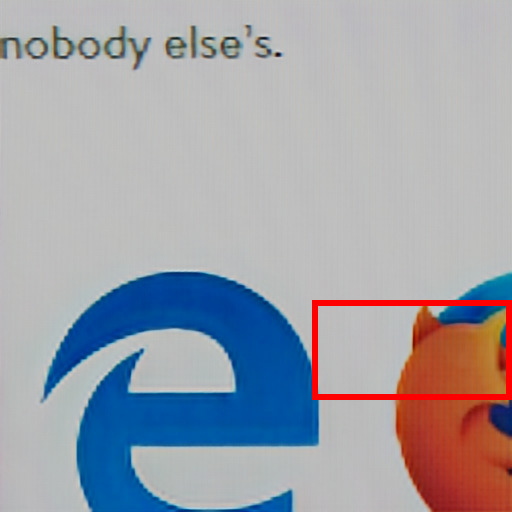}
    \includegraphics[width=\linewidth]{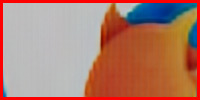}
    \caption{Result of $H'$}
  \end{subfigure}
  \begin{subfigure}{0.32\linewidth}
    \includegraphics[width=\linewidth]{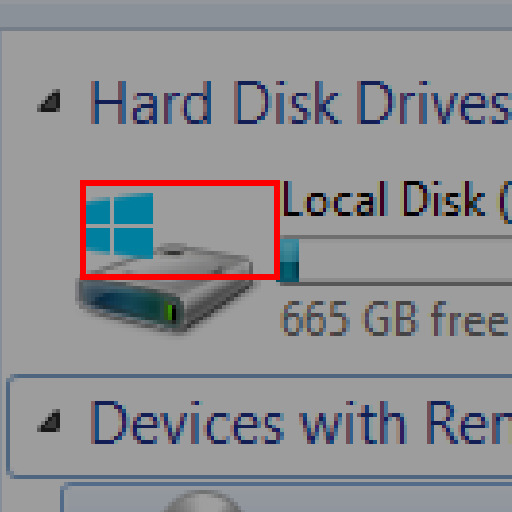}
    \includegraphics[width=\linewidth]{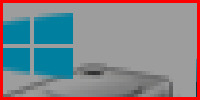}
    \\[1ex]
    \includegraphics[width=\linewidth]{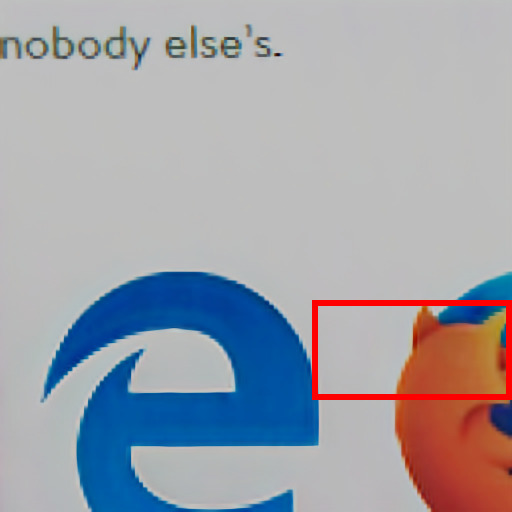}
    \includegraphics[width=\linewidth]{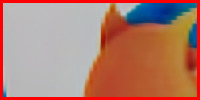}
    \caption{Result of $H$}
  \end{subfigure}

  \caption{For fine-scale demoir\'eing, the real data retrained
    network $H$ can perform better for real images (second row) than
    the synthetic data trained network $H'$.}

  \label{fig:exp_srgan}
\end{figure}

Using the same GAN-based retraining scheme as presented previously, we
can also boost the performance of the fine-scale demoir\'eing network
on real camera-captured screen images.  By retraining using real data,
the improved network $H$ works much better than the synthetic data
trained network $H'$, as demonstrated in the second row of
Fig.~\ref{fig:exp_srgan}.  For synthetic input image, the results of
$H$ and $H'$ are similar, as shown in the first row of
Fig.~\ref{fig:exp_srgan}.

\section{Experiments}
\label{sec:result}

To evaluate the performance of the proposed demoir\'eing algorithm, we
implement the deep convolutional neural network in using Tensorflow.
All of the reported experiments in this paper are conducted on a
computer with a NVIDIA Titan Xp GPU and an Intel i7-4770 CPU.

\subsection{Training Details}
\label{sec:training}

Form 1,000 digital images with various content that is common
displayed on a computer, we create 80,000 $512\times 512$ image
patches with artificial moir\'e patterns using the data synthesizer
presented in Section \ref{sec:data}.  Of the 80,000 patches, 60,000
are used for training and 20,000 are used for testing.  After Gaussian
blurring and downsampling with scale factor 4, each patch is resized
to $128\times 128$ for the training of coarse-scale demoir\'eing
network $G'$.  In the training process, we employ Adam's optimization
method \cite{kinga2015method} with $\beta_1 = 0.9$ and a learning rate
of $10^{-4}$.  The network $G'$ is trained with $2 \times 10^5$ update
iterations, where in each iteration, 32 patches are randomly picked as
a mini-batch.

The synthetic data trained network $G'$ is then used for retraining
GAN-based network $G$.  The input training data for $G$ are real
screen images displayed on 3 different screens (Alienware AW2310, Dell
P2417H, and Samsung SyncMaster T260) and taken by 3 different
smartphones (iPhone 6, iPhone 8, Samsung Galaxy S8).  In total, 300
images with different combinations of devices from different distances
and angles are taken.  From the 300 images, 20,000 $512\times 512$
patches are extracted for our GAN training. 

All the weights in the network $G$ are retrained with $10^4$ update
iterations at a learning rate of $10^{-5}$.  Following the method of
Goodfellow et al. \cite{goodfellow2014generative}, we choose $k = 1$
and alternately update the discriminator and the generator.

For the fine-scale demoir\'eing network $H$, we adopt the same
two-stage training strategy.  First, we pretrain network $H'$ using
synthetic data.  All the input demoir\'ed synthetic images are
upsampled with the Bicubic interpolation method and cropped into
$64\times 64$ patches for the pretraining.  The batch size is 64 and
the number of update iterations is $10^5$.  The Adam optimizer uses
the same setting as in the training of $G$.  After $H'$ being ready,
we retrain the network $H$ using the real camera-captured data for
another $10^4$ iterations at a lower learning rate of $10^{-5}$.

\subsection{Experimental Results}
\label{sec:exp_results}

For performance evaluation, two state-of-the-art general purpose image
restoration algorithms, DnCNN \cite{zhang2017beyond} and RED-Net
\cite{mao2016image} are tested along with the proposed algorithm.
DnCNN can tackle many different image restoration tasks, such as
Gaussian denoising, single image super-resolution and JPEG image
deblocking, and RED-Net is designed to preserve the primary image
components and eliminate various corruptions.

For the demoir\'eing task, we train a 20-layer DnCNN (DnCNN20) with a
receptive field of $41\times 41$ using our synthetic training data
set.  We also employ an extended DnCNN network with 35 layers
(DnCNN35) to achieve a larger receptive field ($71\times 71$) to match
the receptive field of our technique.  For RED-Net, we trained a
20-layer network (RED20) and a 36-layer network (RED36) with default
settings provided by the original authors.  The receptive fields for
RED20 and RED36 are $41\times 41$ and $73\times 73$, respectively.

\begin{table}
  \caption{Average PSNR and SSIM for synthetic data for coarse-scale
    images.}
  \label{tab:psnr_lr}
  \centering
  \resizebox{\linewidth}{!}{
	  \begin{tabular}{cccccc}
	    \hline
	    & DnCNN20 & DnCNN35 & RED20 & RED36 & Ours \\
	    \hline
	    PSNR & 37.26dB & 38.84dB & 38.07dB & 38.53dB & \textbf{41.59dB} \\
	    \hline
	    SSIM & 0.9512 & 0.9858 & 0.9773 & 0.9826 & \textbf{0.9934} \\
	    \hline
	  \end{tabular}
  }
\end{table}

We first test all the algorithms using coarse-scale images, which are
downsampled from the synthetic moir\'e-interfered images.  The peak
signal-to-noise ratio (PSNR) and structural similarity (SSIM) are
calculated for evaluation, as listed in Table~\ref{tab:psnr_lr}.  The
proposed generator achieves much better demoir\'eing performance on
the synthetic images, which exceeds the other compared methods almost
3dB in PSNR and 0.01 in SSIM.

\begin{figure}[t]
  \begin{subfigure}{0.98\linewidth}
    \centering
    \includegraphics[width=0.32\linewidth]{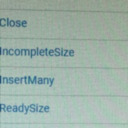}
    \includegraphics[width=0.32\linewidth]{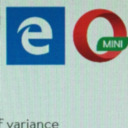}
    \includegraphics[width=0.32\linewidth]{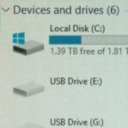}
    \caption{LR Input}
  \end{subfigure}
  \begin{subfigure}{0.98\linewidth}
    \centering
    \includegraphics[width=0.32\linewidth]{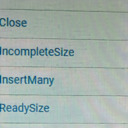}
    \includegraphics[width=0.32\linewidth]{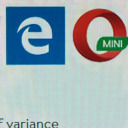}
    \includegraphics[width=0.32\linewidth]{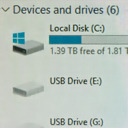}
    \caption{DnCNN35}
  \end{subfigure}
  \begin{subfigure}{0.98\linewidth}
    \centering
    \includegraphics[width=0.32\linewidth]{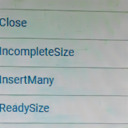}
    \includegraphics[width=0.32\linewidth]{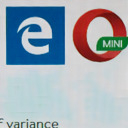}
    \includegraphics[width=0.32\linewidth]{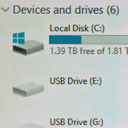}
    \caption{RED36}
  \end{subfigure}
  \begin{subfigure}{0.98\linewidth}
    \centering
    \includegraphics[width=0.32\linewidth]{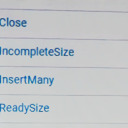}
    \includegraphics[width=0.32\linewidth]{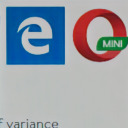}
    \includegraphics[width=0.32\linewidth]{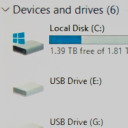}
    \caption{The proposed}
  \end{subfigure}
  \caption{Results of the tested demoir\'eing algorithms on
    downsampled camera-captured screen images.}
  \label{fig:lr_exp}
\end{figure}

We also test the trained models on the downsampled real
camera-captured screen images.  As shown in Fig.~\ref{fig:lr_exp}, the
proposed coarse-scale network works well against different
coarse-scale moir\'e patterns.  The other tested general restoration
methods, on the other hand, all fail to remove those wide moir\'e
color bands, despite having similar receptive field as the proposed
technique.

\begin{table}[t]
  \centering
  \caption{Average PSNR and SSIM for synthetic data in high-resolution space.}
  \label{tab:psnr}
  \resizebox{\linewidth}{!}{
	  \begin{tabular}{cccccc}
	    \hline
	    & DnCNN20 & DnCNN35 & RED20 & RED36 & Ours \\
	    \hline
	    PSNR & 35.61dB & 37.46dB & 37.68dB & 37.80dB & \textbf{40.01dB} \\
	    \hline
	    SSIM & 0.9171 & 0.9678 & 0.9698 & 0.9717 & \textbf{0.9829} \\
	    \hline
	  \end{tabular}
  }
\end{table}

\begin{figure}[h]
	\centering
	\includegraphics[width=\linewidth]{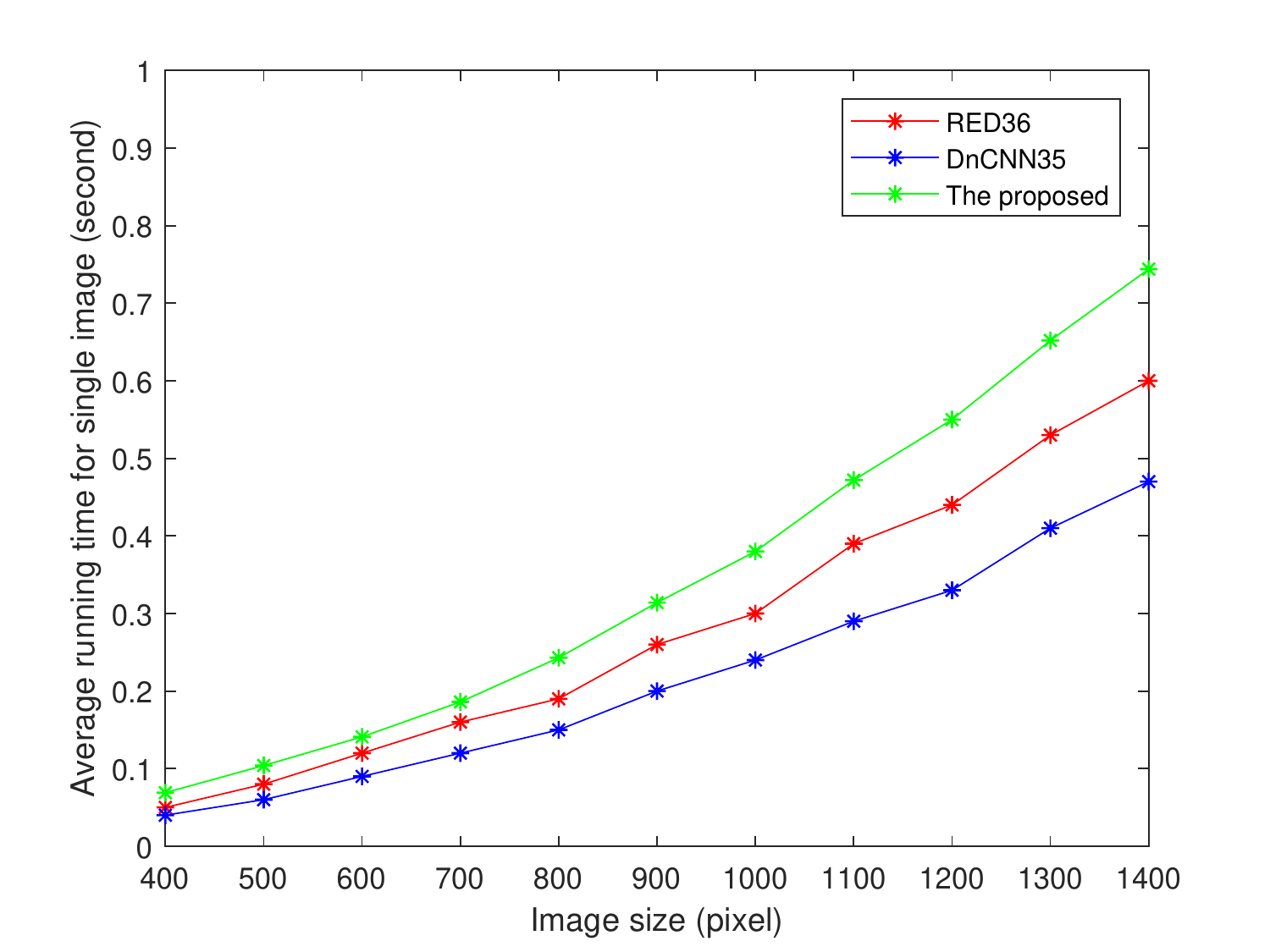}
	\caption{Timing on GPU for all the compared method.}
	\label{fig:timing}
\end{figure}

\begin{figure*}
  \centering
  \includegraphics[width=0.23\linewidth]{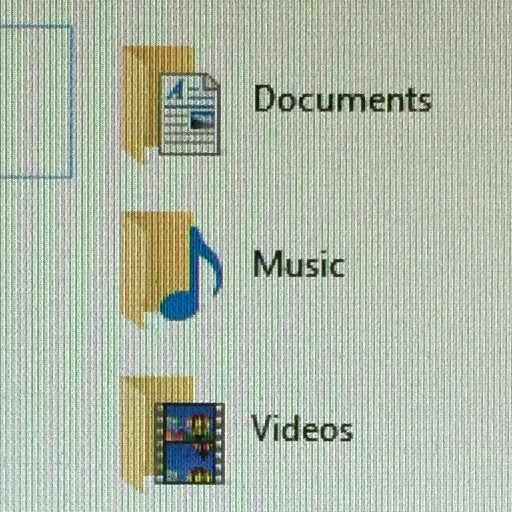}
  \includegraphics[width=0.23\linewidth]{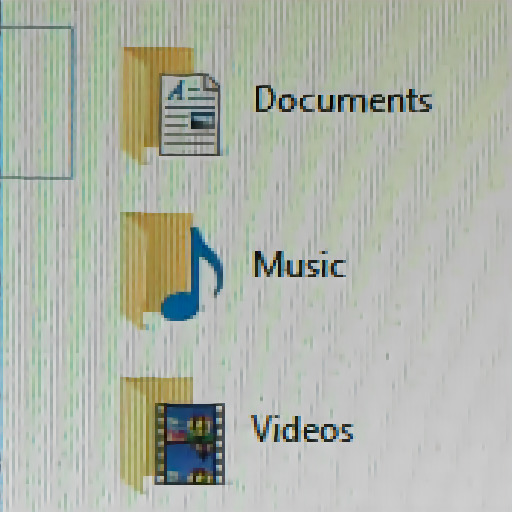}
  \includegraphics[width=0.23\linewidth]{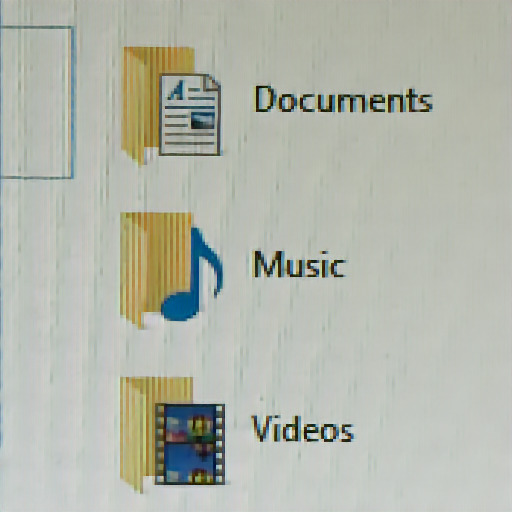}
  \includegraphics[width=0.23\linewidth]{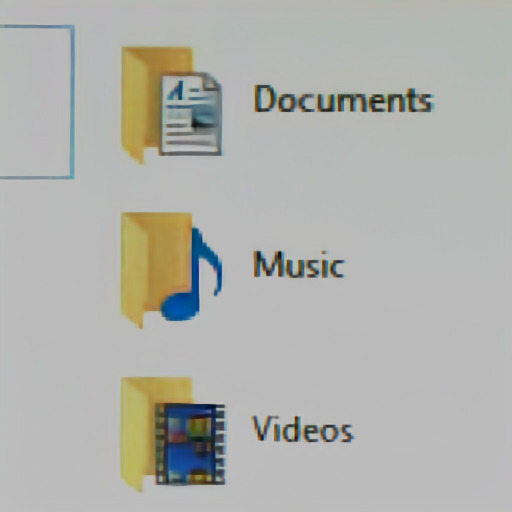} \\
  \vspace{0.2ex}
  \includegraphics[width=0.23\linewidth]{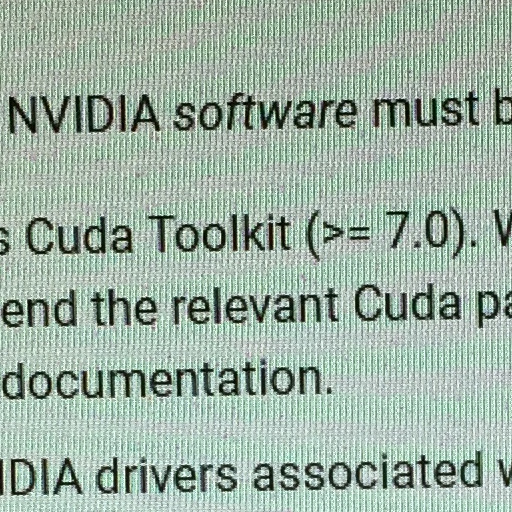}
  \includegraphics[width=0.23\linewidth]{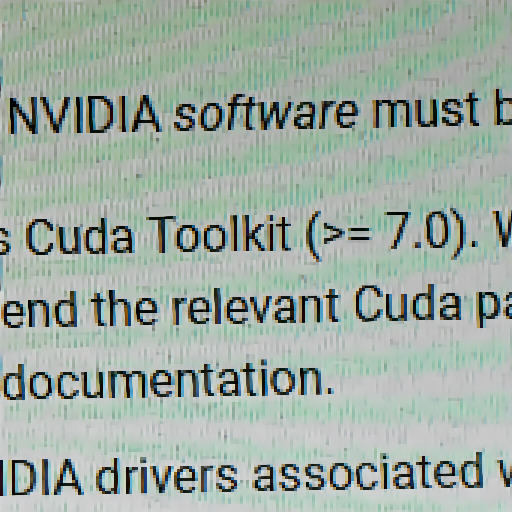}
  \includegraphics[width=0.23\linewidth]{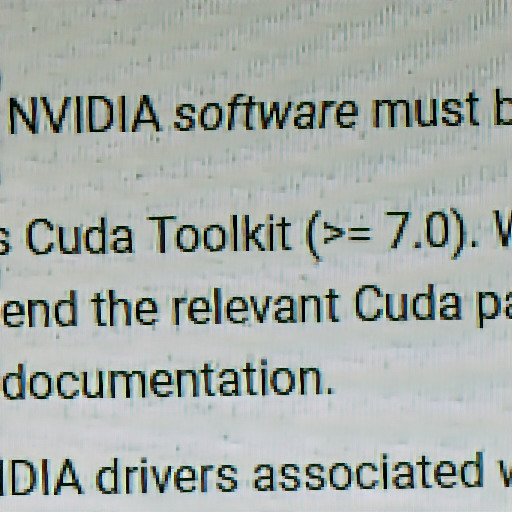}
  \includegraphics[width=0.23\linewidth]{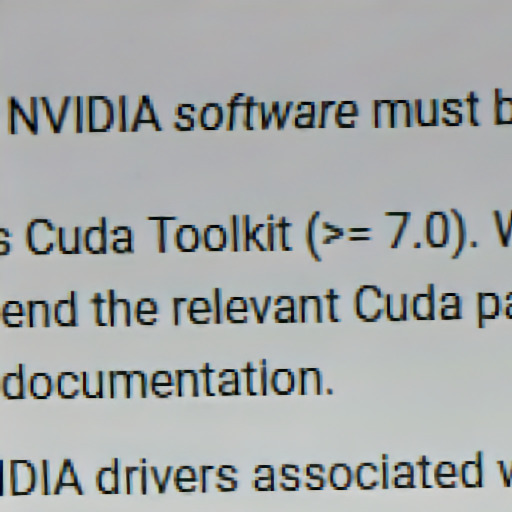} \\
  \vspace{0.2ex}
  \includegraphics[width=0.23\linewidth]{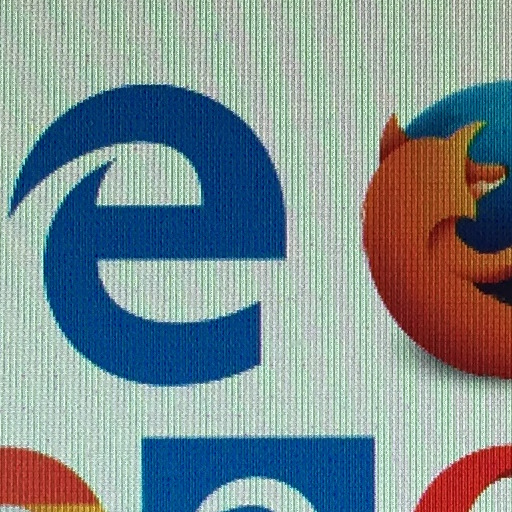}
  \includegraphics[width=0.23\linewidth]{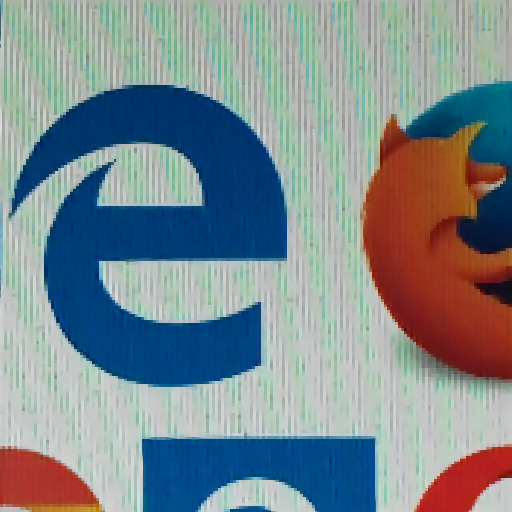}
  \includegraphics[width=0.23\linewidth]{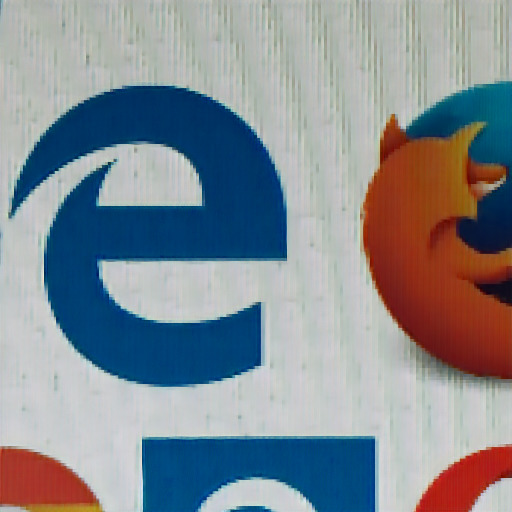}
  \includegraphics[width=0.23\linewidth]{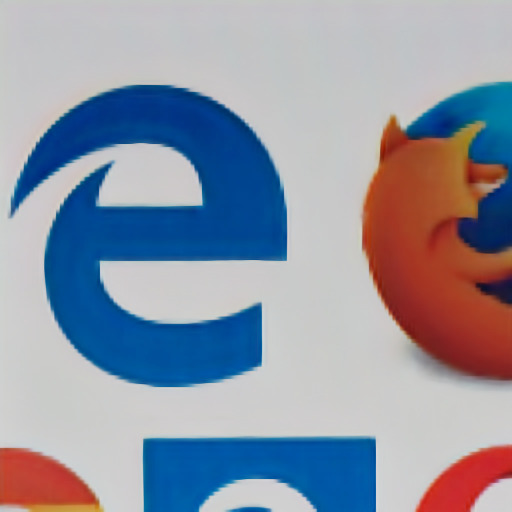} \\
  \vspace{0.2ex}
  \includegraphics[width=0.23\linewidth]{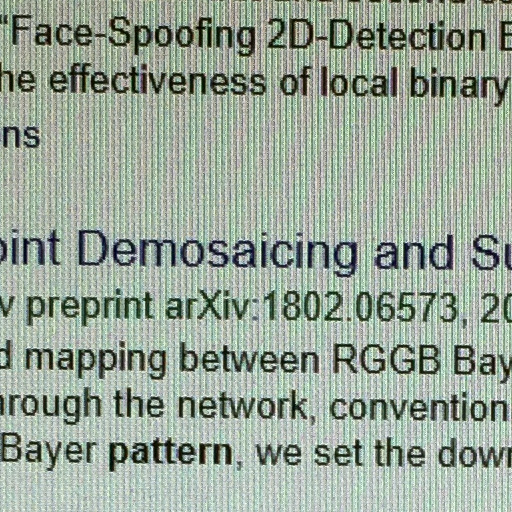}
  \includegraphics[width=0.23\linewidth]{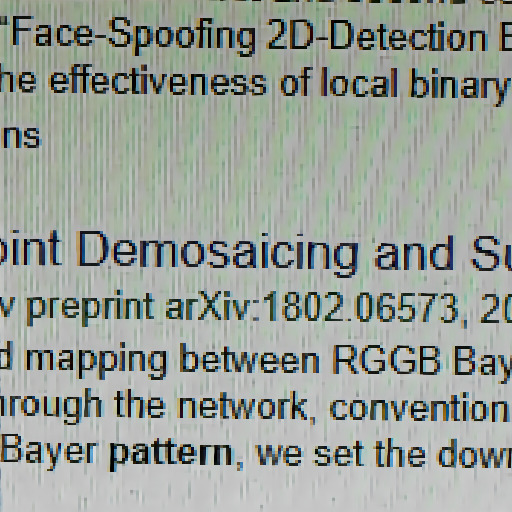}
  \includegraphics[width=0.23\linewidth]{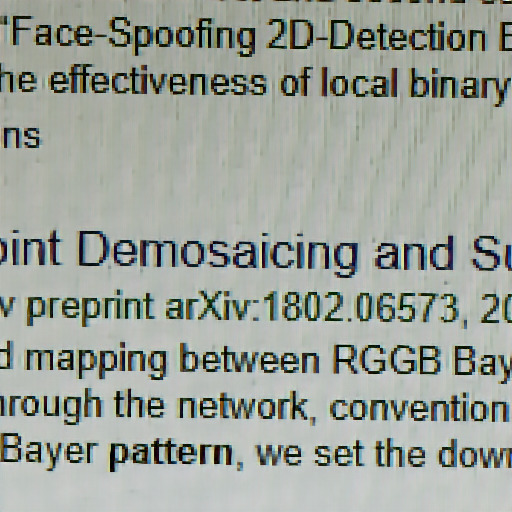}
  \includegraphics[width=0.23\linewidth]{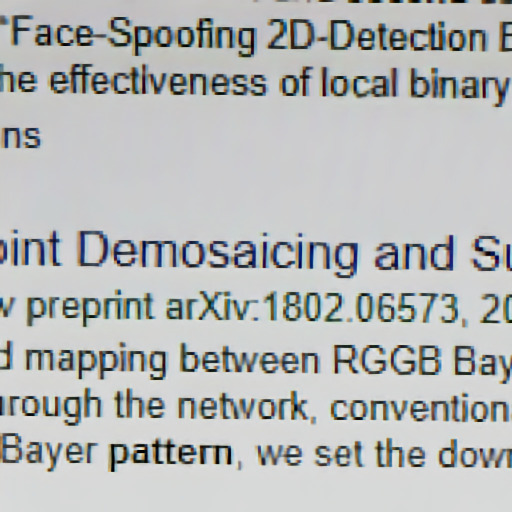} \\
  \vspace{0.2ex}
  \begin{subfigure}{0.23\linewidth}
    \centering
    \includegraphics[width=\textwidth]{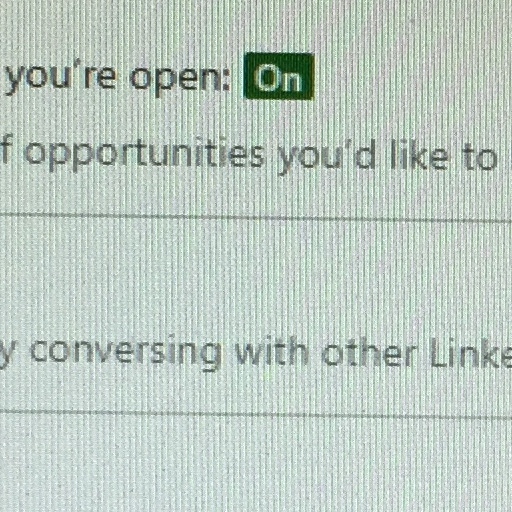}
    \caption{Input}
  \end{subfigure}
  \begin{subfigure}{0.23\linewidth}
    \centering
    \includegraphics[width=\textwidth]{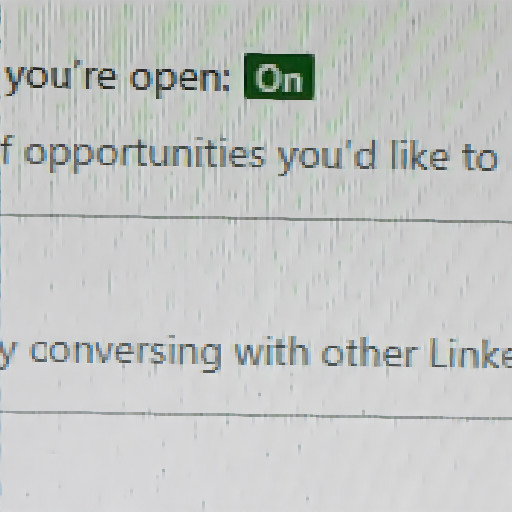}
    \caption{RED36}
  \end{subfigure}
  \begin{subfigure}{0.23\linewidth}
    \centering
    \includegraphics[width=\textwidth]{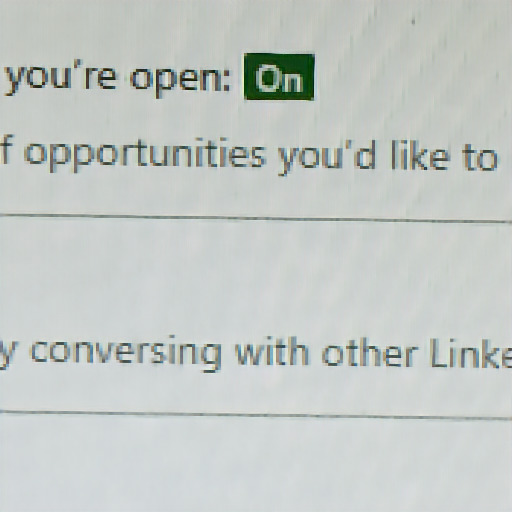}
    \caption{DnCNN35}
  \end{subfigure}
  \begin{subfigure}{0.23\linewidth}
    \centering
    \includegraphics[width=\textwidth]{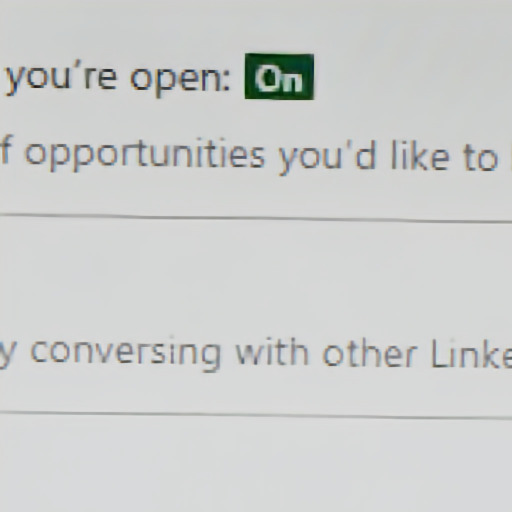}
    \caption{The proposed}
  \end{subfigure}
  \caption{Samples of demoir\'eing results by the tested techniques
    using real fine-scale camera-captured screen images.}
  \label{fig:exp_res}
\end{figure*}

For fine-scale demoir\'eing, we retrain all the DnCNN and RED-Net
models using the synthetic data of original resolutions.  As shown in
Table~\ref{tab:psnr}, the proposed technique is still substantially
better than the competition in terms of PSNR and SSIM.  The advantage
of the proposed technique is also evident for human viewers.  As shown
in Fig.~\ref{fig:exp_res}, the results of the proposed technique are
much cleaner than the other tested methods.

Plotted in Fig.~\ref{fig:timing} is the running time of the tested
algorithms as a function of the size of the input images.  Benefiting
from the multi-scale DCNN design and two-stage training strategy, the
proposed algorithm presents much better demoir\'eing results without
significantly increasing the computational cost.

\section{Conclusion}
\label{sec:conclusion}

Based on the observation that Moir\'e patterns exist and exhibit
vastly different characteristics in different scales, we purposefully
designed a coarse-to-fine DCNN technique for the task of demoir\'eing.
By incorporating a novel retraining strategy, the proposed technique
can work well for real camera-captured screen images even without
paired real images for training.  Extensive experimental results have
demonstrated that the proposed technique can efficiently remove the
moir\'e patterns for camera acquired screen images; the new technique
outperforms the existing ones.

\bibliographystyle{ieee}
\bibliography{demoire}

\end{document}